\documentclass[twocolumn,showpacs,preprintnumbers,groupedaddress,amsmath,amssymb,floatfix,prl]{revtex4}
\usepackage{graphicx,color,epsfig}
\usepackage{ulem}
\begin{document}

\title{Ultracold Plasma Expansion in a Magnetic Field}

\author{X. L. Zhang, R. S. Fletcher, and S. L. Rolston}
\affiliation{Joint Quantum Institute, Department of Physics, University of Maryland, College Park, MD 20742}
\author{P. N. Guzdar, and M. Swisdak}
\affiliation{Institute for Research in Electronics and Applied Physics, University of Maryland, College Park, MD 20770}
\date{\today}

\begin{abstract}
  We measure the expansion of an ultracold plasma across the field lines of a uniform magnetic field.  We image the ion distribution by extracting the ions with a high voltage pulse onto a position-sensitive detector. Early in the lifetime of the plasma ($< 20$ $\mu$s), the size of the image is dominated by the time-of-flight Coulomb explosion of the dense ion cloud. For later times, we measure the 2-D Gaussian width of the ion image, obtaining the transverse expansion velocity as a function of magnetic field (up to 70 G). We observe that the expansion velocity scales as B$^{-1/2}$, explained by a nonlinear ambipolar diffusion model with anisotropic diffusion in two different directions. 
\end{abstract}

\pacs{ 52.25.Xz, 52.55.Dy}
\maketitle
 
Plasma expansion in a uniform magnetic field is of interest in astrophysical, ionospheric, and laser-produced plasma applications. The presence of a magnetic field during expansion can initiate various phenomena, such as plasma confinement and plasma instabilities \cite{harilal2004}. Ultracold plasmas (UCPs), formed by photoionizing laser-cooled atoms near the ionization limit, have system parameters many orders of magnitude away from traditional laser-produced plasmas, with electron and ion temperatures  on the order of meV, or even $\mu$eV, and densities of $10^5$ to $10^{10}$ ${\rm cm}^{-3}$. UCPs thus provide a testing ground to study basic plasma theory in a clean and simple system, and we need fields of only tens of Gauss  to observe significant effects on the expansion dynamics (for our system the electrons are magnetized, while the ions are not). All previous studies of ultracold plasma expansion, both experimental \cite{kulin2000, simien2004, cummings2005, fletcher2006} and theoretical \cite{bergeson2003, robicheaux2003, pohl2004, mazevet2002}, focus on free expansion in the absence of magnetic fields.

\begin{figure}[htbp]
\begin{center}
\epsfig{file=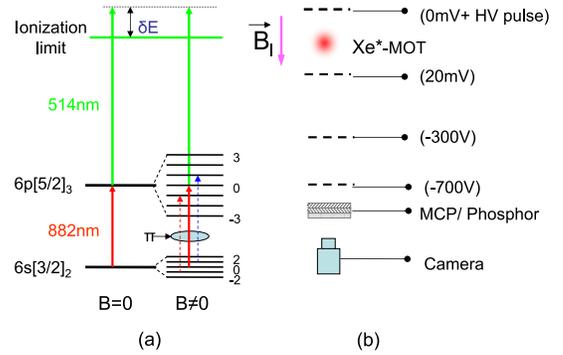, width=2.80in}
\end{center}
\caption{ Experimental setup and two-photon excitation process for an UCP in a magnetic field. (a) two-photon excitation process: one photon (red solid line, $\pi$ polarized)  at 882 nm drives the $6s [3/2]_2$ (m = 0) $\rightarrow$  $6p [5/2]_3$ (m = 0) transition, and the other (10-ns pulse) at 514 nm ionizes the atoms in the $6p [5/2]_3$ (m = 0) state. The dashed lines are the repumping beams (both blue and red of the $6s [3/2]_2$ $\rightarrow$  $6p [5/2]_3$ transition, also $\pi$ polarized); (b) experimental setup for imaging the ions onto the MCP/phosphor screen.}
\end{figure}
In this work, we present the first measurement of UCP expansion in a magnetic field. By extracting the ions with a high voltage pulse onto a position-sensitive detector, we image the ion distribution of an UCP expanding in a uniform  magnetic field. Early in the lifetime of the plasma, the  image size is dominated by the time-of-flight Coulomb explosion of the dense ion cloud. At about 20 $\mu$s the image size is at a minimum and then afterwards increases, reflecting the true size of the expanding plasma. The expansion is self-similar, as the ion cloud maintains a Gaussian density profile throughout the lifetime of the plasma. By 2-D Gaussian fitting of the ion image, we obtain the width transverse to the applied magnetic field. In the absence of a magnetic field, the plasma expansion velocity at different initial electron temperatures ($T_e$) matches the result obtained by measuring the plasma oscillation frequency \cite{kulin2000, zhang2007}. As we increase the field up to 70 G, we find that the transverse expansion velocity decreases, roughly scaling as B$^{-1/2}$. This field dependence is well explained by an ambipolar diffusion model which involves anisotropic diffusion in two different directions: the diffusion rate is almost unaffected in the direction along the magnetic field, while it is reduced in the direction normal to the field.  A critical feature of this diffusion model is its dependence on density and temperature which, for an expanding UCP, are time dependent.

Our production of UCPs proceeds as detailed in  \cite{killian1999}. Metastable Xenon atoms are cooled and trapped in a magneto-optical trap (MOT). The neutral atom cloud has a temperature of about 20 $\mu$K, total number of about $2 \times 10^6$, peak density of about 2 x 10$^{10}$ cm$^{-3}$, and a Gaussian spatial density distribution with an rms radius of about 280 $\mu$m. The plasma is then produced by a two-photon excitation process (figure 1a), ionizing up to 30$\%$ of the atoms. One photon for this process is from the cooling laser at 882 nm, and the other is from a pulsed dye laser at 514 nm (10-ns pulse). We control the ionization fraction with the 514-nm intensity, while the initial electron energy is controlled by tuning the 514-nm photon energy with respect to the ionization limit. In the absence of a magnetic field, we can simply use the MOT beams to drive the $6s [3/2]_2$ $\rightarrow$ $6p [5/2]_3$ transition; with the addition of a magnetic field, the $6s [3/2]_2$ and $6p [5/2]_3$ states split into 5 and 7 Zeeman sub-levels respectively, shifting the transitions out of resonance. To maintain a reasonable excitation fraction, we need to optically pump the atoms into the m = 0 magnetic sublevel of the ground state and then drive the  $6s [3/2]_2$  (m = 0) $\rightarrow$  $6p [5/2]_3$ (m = 0) transition, which will be unaffected by the magnetic field to first order. (We cannot turn on magnetic fields in a time short compared to the time that ions acquire velocity in the plasma due to electron pressure ($\sim$ 1 $\mu$s)  and thus must photoionize in the presence of the field). It is difficult to accumulate a large m = 0 population via optical pumping for a $J \rightarrow J+1$ transition, so we developed a multiple frequency scheme that uses Zeeman shifts to address the m = 0 and m = $\pm$ 1 sublevels (figure 1a).  We use frequencies detuned -5, 0, 5 MHz respectively relative to the $6s [3/2]_2$  (m = 0) $\rightarrow$  $6p [5/2]_3$ (m = 0) transition, and generate three $\pi$-polarized optical pumping beams. With this configuration, we can achieve UCP densities of $\sim$ 50\% of the zero field value, approximately independent of magnetic field up to 100 G.

\begin{figure}[htbp]
\begin{center}
\epsfig{file=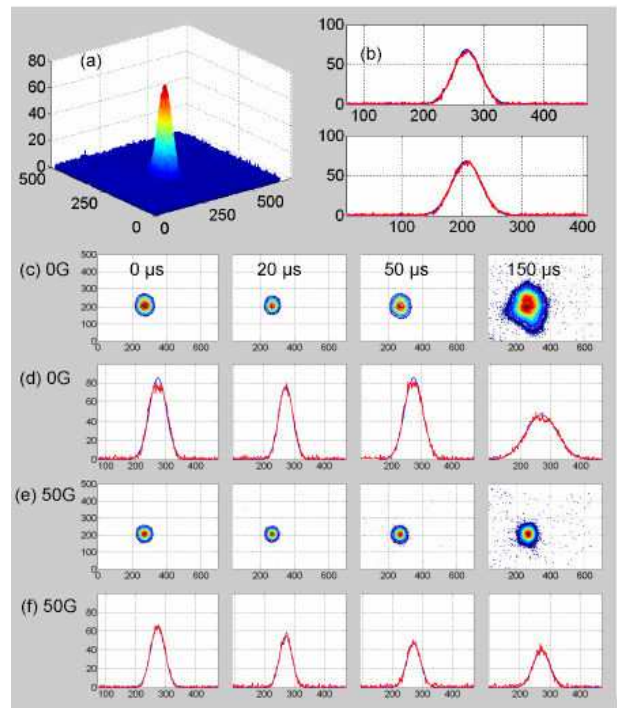, width=3.20in}
\end{center}
\caption{(a) a false color 2D ion spatial distribution integrated over the third dimension of the UCP for t = 20 $\mu$s, $T_e$ = 100 K and B = 20 G; (b) the 2D Gaussian fittings (blue curves) of the ion image (a) along the x and y axis in the horizontal plane (red curves);(c) and (e) are the contour plots of the ion images at different delay times for 0 G and 50 G respectively; (d) and (f) are the corresponding 2-D Gaussian fit of (c) and (e). All the units in (a)-(f) are in pixel number, and 1 pixel unit is about 150 $\mu$m.}
\end{figure}

The ionized cloud rapidly loses a few percent of the electrons, resulting in a slightly attractive potential for the remaining electrons, and it quickly reaches a quasineutral plasma state. It then expands with an asymptotic velocity typically in the 50-100 m/s range caused by the outward electron pressure \cite{kulin2000}. By applying a high-voltage pulse to the top grid (about 1.5 cm above the plasma) and with accelerating voltages on the middle and front grids (figure 1b), we image (magnification of 1.3) the ion distribution of the UCP on a position-sensitive microchannel plate detector with phosphor screen.  The phosphor images are recorded with a CCD camera.  The high-voltage pulse has an amplitude of 340 V, a width of 4.5 $\mu$s, and a rise time of 60 ns. We use a modified square pulse generator  \cite{tomic1990}, which uses  power FETs to fast switch a high voltage source. The magnetic field is generated by a pair of coils which are located outside the vacuum chamber, and is fully turned on before the creation of the UCP. Figure 2 is a typical result with an average of 8 images to increase the signal-to-noise ratio. Figure 2a is a false color plot of an ion image after 20 $\mu$s of plasma evolution for $T_{e }(0)$ = 100 K and B = 20 G, which fits well to a  2-D Gaussian profile (figure 2b). The ion images maintain a Gaussian profile during the whole lifetime of the UCP (about 200 $\mu$s) as shown in figure 2 c-f. By 2-D Gaussian fitting of the ion image, we extract the size  transverse to the magnetic field at specific delay times. In figure 3, we plot the transverse sizes of UCPs in a field as a function of time. The measured size of the image for early times is dominated by the Coulomb explosion of the ions, and reflects the overall Coulomb energy rather than the plasma size \cite{zhang2007, CoulombEffect}. As the plasma size increases, the Coulomb explosion effect diminishes (note decreasing images at early times). After about 20 $\mu$s, the measured image reflects the plasma size, and the images increase with increasing plasma time, as expected. As we increase the field, the transverse size increases more slowly with time, i.e., the slopes of the curves decrease, which indicates a slower expansion and magnetic confinement in the transverse direction of the UCP.  For 0 G at long times, the size does not linearly increase. This is in part because the size of the UCP is large enough to be affected by the four posts that secure our grids above and below the plasma. We also notice that the ion images have a flat top and even dip at late times of about 150-200 $\mu$s for the 0 G case, the cause of which is unknown.   

\begin{figure}[htbp]
\begin{center}
\epsfig{file=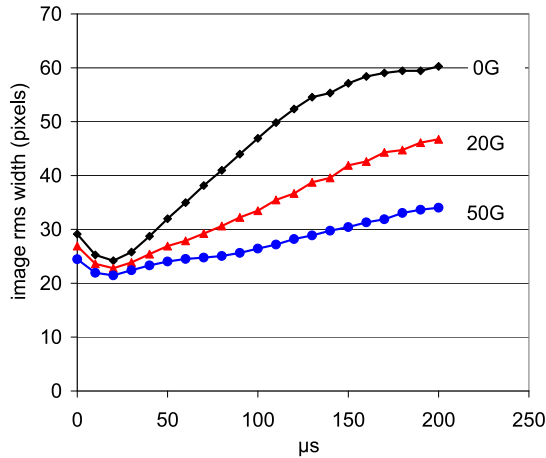, width=2.80in}
\end{center}
\caption{ Transverse sizes of UCPs in a magnetic field. The black curve (diamonds) is for 0 G, the red curve (triangles) is for 20 G, and the blue curve (dots) is for 50 G. Early in the lifetime of the plasma, the size of the image is dominated by the Coulomb explosion of the dense ion cloud.}
\end{figure}

By linear fitting to the transverse sizes vs. time after about 20 $\mu$s (for 0 G, only fitting the restricted linear region), we extract the transverse asymptotic expansion velocity of the UCP. In figure 4, we plot the transverse asymptotic expansion velocity as a function of magnetic field. The blue solid curve with square points is the experimental result, the black dashed curve is a power-law fit of the data, and the red curve with triangle points is a simulation by assuming that the transverse expansion is due to ambipolar diffusion (discussed below). In the absence of a magnetic field and $T_{e} (0)$ = 100 K, the UCP expands with an asymptotic velocity of about 70 m/s caused by the outward electron pressure. This is the ion acoustic velocity for $T_e (0)$.  As the field increases, the transverse expansion velocity decreases, roughly scaling as B$^{-1/2}$. At 50 G, it is about 17 m/s, about a factor of four smaller than that at 0 G.

Using the same technique, we extract the transverse asymptotic expansion velocity for different $T_e (0)$s, shown in fig. 5. The red solid line with square points is the experimental result for 0 G, which agrees with the results obtained by measuring the plasma oscillation frequency (the black solid curve with circle points) \cite{kulin2000}. This validates our imaging method for measuring the transverse expansion velocity and previous thechnique. The dashed lines are linear fits of the data above 60 K. For 50 G, the transverse expansion velocity as a function of $T_{e} (0)$ is similar to that at 0 G except for a factor 3-4 smaller. The expansion velocity goes as ${T_e}^{1/2}$ for initial electron temperature higher than 60 K, that is, the slopes of the dashed lines in figure 5 are about 1/2.

\begin{figure}[htbp]
\begin{center}
\epsfig{file=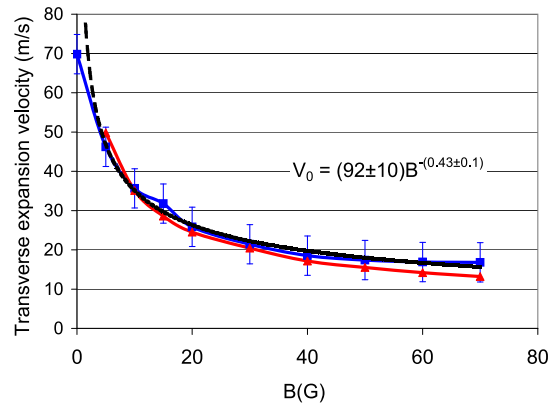, width=2.80in}
\end{center}
\caption{ The transverse asymptotic expansion velocity vs. applied magnetic field. The blue curve (squares) is the experimental result, the black dashed curve is a power-law fit, and the red curve (triangles) is a simulation using an ambipolar diffusion model that involves anisotropic diffusion in two directions.The error bars represent the 1$\sigma$ standard uncertainty resulting from the linear fits to the curves in figure 3. }
\end{figure}
 
\begin{figure}[htbp]
\begin{center}
\epsfig{file=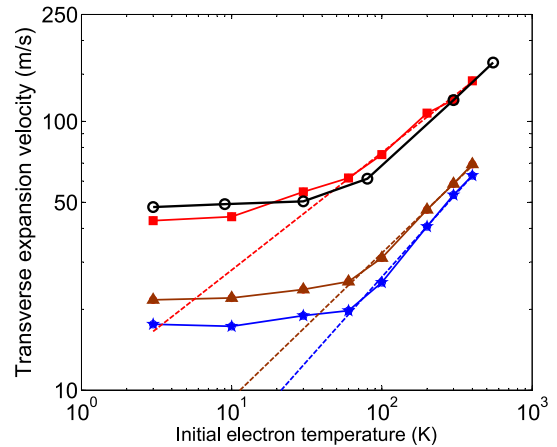, width=2.80in}
\end{center}
\caption{ The transverse expansion velocity as a function of initial electron temperature $T_{e}(0)$. The red solid curve (squares) is the experimental result for 0 G, which matches results obtained by measuring the plasma oscillation frequency (the black solid curve (circles)) \cite{kulin2000}. The brown solid curve (triangles) is for 25 G, and the blue solid curve (stars) is for 50 G. The dashed lines are power-law fits  above 60 K with powers near 1/2 (0.44, 0.54 and 0.63 from top to bottom, respectively).} 
\end{figure}  

  The expansion of the UCP at the ion acoustic velocity can be understood in terms of electron pressure in a simple hydrodynamic model \cite{kulin2000}, or alternatively as ambipolar diffusion, where oppositely charged particles (electrons and ions) diffuse together at the lower ion velocity. In a magnetic field, ambipolar diffusion allows charged particles to cross field lines, and produces a diffusion constant that scales as $B^{-2}$, and a transverse decay size (expansion velocity) scaling of $B^{-1}$  for conventional low temperature plasmas such as arc plasmas and strongly ionized plasmas in steady state (Q machine) \cite{hoh1962}, in contrast to our observed $B^{-1/2}$ scaling.  Another possibility is anomalous diffusion (Bohm diffusion) with a diffusion constant scaling as $B^{-1}$, which is intimately connected to plasma instabilities \cite{boeschoten1964, hoh1962}. However, observing the electron signals as well as the single shot ion images with the magnetic fields, we do not see any signature of plasma instabilities.
Our UCPs are rather novel plasma systems, not only because of their low temperatures, but also because they are unconfined, inhomogeneous, and freely expanding into vacuum (the same properties  hold for the UCP expansion along the magnetic field). There is also competition between adiabatic cooling and three-body recombination-induced heating leading to a time-dependent falling $T_{e}$ \cite{fletcher2007}. Given the evolution of the plasma, we cannot simply apply steady-state ambipolar diffusion theory, but must include the dynamics. In a strong magnetic field, the diffusion rate of a plasma becomes anisotropic; the diffusion rate is almost unaffected in the direction along the magnetic field, while it is reduced in the direction normal to the field. For our UCP expansion in the magnetic field, we assume ambipolar diffusion in both directions (along the field and normal to the field) and write the diffusion equation as follows by separating the diffusion rate into two terms in cylindrical coordinates:
\begin{equation}
\frac {\partial n}{\partial t} =  \frac {1}{r} \frac{\partial}{\partial r}rD_\bot \frac {\partial n}{\partial r} -\frac {n}{\tau_\parallel} 
\end{equation}
where $D_\bot$ is the transverse ambipolar diffusion rate with $D_\bot={{\rho}^2_e}\nu_{ei}\propto \frac {n_e}{B^2{T_e}^{1/2}}$, $\rho_e$ is the electron gyroradius and $\nu_{ei}$ is the electron-ion collision rate. $\tau_{\parallel}$ is the characteristic time with $\tau_{\parallel}=a(t)/v_s(t)$, a(t)  and $v_s(t)$ are the plasma size and the expansion velocity along the magnetic field respectively. The first term in the right hand side of eq.1 corresponds to the diffusion normal to the field, and the second term is the diffusion along the magnetic field. Normalizing all the variables in eq. 1 to their initial values, we get:
\begin{equation}
\frac {\partial{\tilde{n}}}{\partial {\tilde{t}}} =  \frac {1}{\tilde{r}}\frac {\partial}{\partial {\tilde{r}}}\frac {\tilde{r}\tilde{n}}{{\tilde{T}}^{1/2}}\frac {\partial \tilde{n}}{\partial \tilde{r}} -\frac {\alpha \tilde{n}}{\tilde{\tau}_{\parallel}} 
\end{equation}
where $\tilde{n}=\frac {n}{n_0}$, $\tilde{t}=\frac {t}{t_0}$, $\tilde{r}=\frac {r}{a_0}$, $\tilde{T}=\frac {T}{T_0}$, $\tilde{\tau}_{\parallel}=\frac {\tau_{\parallel}}{\tau_{\parallel 0}}$, $\alpha=\frac {t_0}{\tau_{\parallel 0}} $, with $t_0=\frac {a^2_0}{{{\rho}^2_{e0}}\nu_{ei0}}$ and $\tau_{\parallel 0}=\frac {a_0} {v_{s0}}$.

For simplicity, we assume no spatial temperature gradient for our UCP, and that the density profiles in both directions remain Gaussian throughout the plasma lifetime. In order to solve eq. 2, we also need to know $T_{e} (t)$  in the magnetic field. Although we might expect changes in the $T_{e}$ evolution in the presence of a large magnetic field compared with free expansion, we did not observe any significant difference in the three-body recombination rates, which can be used to extract $T_{e} (t)$ \cite{ fletcher2007}.  We therefore assume that $T_{e} (t)$ is independent of magnetic field for the sake of our diffusion model (eq. 2). Electron temperature evolution in an expanding UCP is a complicated problem and has been studied using various methods \cite{roberts2004, fletcher2006, gupta2007, fletcher2007}. Using Doppler broadening of the ion optical absorption spectrum together with numerical simulation, $T_{e} (t)$ was observed to follow a self-similar analytic solution during the first 20 $\mu$s for $T_{e} (0)$ greater than 45K, that is $T_e = T_e(0)/(1+t^2/{\tau}^2_{\parallel 0})$ \cite{gupta2007}. We obtain $T_{e} (t)$ up to 70 $\mu$s by using three-body recombination rate \cite{fletcher2007}, which gives $T_e \propto t^{-1.2}$ for $T_{e0} = 3 K$. We use the following empirical equation for  $T_{e} (t)$ for our diffusion model:
\begin{eqnarray}
T_e &=& T_e(0)/(1+t^2/{\tau}^2_{\parallel 0}),t< 20 \mu s \nonumber \\
T_e &=& T_e(20 \mu s) (t/20)^{-1.2}, t \geq 20 \mu s.
\end{eqnarray}

By numerically solving eq. 2 with $T_{e} (t)$ given by eq. 3, we find the transverse size evolution for different magnetic fields from 5 G to 70 G. We extract the transverse expansion velocity as a function of magnetic field, shown as the red curve (triangles) in figure 4. The simulation result is in good agreement with the $B^{-1/2}$ scaling observed in the experimental data. Therefore ambipolar diffusion can explain the B-dependence when the expansion dynamics are included. Using self-similiar arguments for the diffusion \cite{barenblatt1978}, it can be shown that $r\propto B^{-1/2}t^{(1+\Delta)/4}$ where $T_e\propto t^{-\Delta}$. By taking the derivative of r, we find an expansion velocity scaling as $B^{-1/2}$, as observed. Even if $T_{e} (t)$ in the presence of magnetic field is different from eq. 3 (different $\Delta$), it will only alter the time-dependence of the transverse size, but not the $B^{-1/2}$ scaling of the above self-similiar argument. This scaling of the transverse size (expansion velocity) is the result of the electron density dependence of the tranverse ambipolar diffusion rate, while that of other low temperature plasmas is density independent and yields the usual $B^{-1}$ dependence of the characteristic transverse decay distance. Bohm diffusion also has the same $B^{-1/2}$ dependence of the transverse size for our UCP systems if we assume $D_\bot \propto {T_e}/{eB}$. However, the plasma will quickly stop expanding in a very short time ($<$ 20 $\mu$s) if we assume only Bohm diffusion is involved, because the Bohm diffusion rate is proportional to $T_e$. The ambipolar diffusion rate is inversely proportional to ${T_e}^{1/2}$, and should dominate Bohm diffusion at later times which is consistent with no signature of plasma instabilities.

In conclusion, we have developed a method to study the transverse expansion of an ultracold plasma in a magnetic field by using a high voltage pulse to image the ion distribution onto a position sensitive detector. The expansion velocity roughly scales as $B^{-1/2}$ for magnetic field up to 70 G, and is in good agreement with a nonlinear ambipolar diffusion model.

\begin{acknowledgments}
This work was partially supported by the National Science Foundation PHY-0245023 and PHY-0714381.
\end{acknowledgments}


\begin{thebibliography}{99}
\bibitem{CoulombEffect} The electrons are quickly removed from the UCP in a few ns by the HV pulse, but the ions maintain their spatial distribution. The strong Coulomb forces between the ions (due to the small ion cloud size in the first 20 $\mu$s) produce images proportional to the Coulomb energy of the ion cloud.
\end{thebibliography}

\end{document}